\documentclass[AMA]{WileyNJD-v1}
\usepackage{graphicx}
\usepackage{geometry}
\usepackage{float}
\usepackage{amsmath}
\usepackage{multirow}
\usepackage{amssymb}
\usepackage{caption}
\usepackage{amssymb}
\usepackage{amsthm}
\usepackage{palatino}
\usepackage{listings,xcolor}
\articletype{Research Paper}%
\begin{document}

\title{Modeling Coefficient Alpha for Measurement of Individualized Test Score Internal Consistency}

\author[1]{Molei Liu}

\author[2]{Ming Hu}

\author[3]{Xiaohua Zhou*}

\authormark{Liu \textsc{et al}}

\address[1]{\orgdiv{Department of Probability and Statistics, School of Mathematical Sciences}, \orgname{Peking University}, \orgaddress{\state{Beijing}, \country{China}}}

\address[2]{\orgdiv{Department of Epidemiology and Biostatistics, School of Public Health}, \orgname{Central South University}, \orgaddress{\state{Hunan}, \country{China}}}

\address[3]{\orgdiv{Beijing International Center for Mathematical Research}, \orgname{Peking University}, \orgaddress{\state{Beijing}, \country{China}}}

\corres{Xiaohua Zhou, *\email{azhou@bicmr.pku.edu.cn}}

\presentaddress{Room 78 Jingchunyuan, Peking University, No.5 Yiheyuan Road, Haidian District, Beijing, China.}

\abstract[Summary]{A method for measuring individualized reliability of several tests on subjects with heterogenecity is proposed. A regression model is developed based on three sets of generalized estimating equations (GEE). The first set of GEE models the expectation of the responses, the second set of GEE models the response's variance, and the third set is proposed to estimate the individualized coefficient alpha, defined and used to measure individualized internal consistency of the responses. We also extend our method to handle missing data in the covariates. Asymptotic property of the estimators is discussed, based on which interval estimation of the coefficient alpha and significance detection are derived. Performance of our method is evaluated through simulation study and real data analysis. The real data application is from a health literacy study in Hunan province of China.}

\keywords{reliability, coefficient alpha, missing data, generalized estimating equation, asymptotic normality, confidence interval, hypothesis testing.}

\maketitle

\section{Introduction}
For tests and questionnaires with multiple items, reliability is a fundamental elements in the evaluation of the measurement quality. Coefficient alpha (or Cronbach's alpha) was proposed by (Cronbach, 1951\cite{Cronbach1951Coefficient}, 1988\cite{Cronbach1988Internal}; Cronbach and Shavelson, 2004\cite{Cronbach2004My}), and has been widely used in social, behavioral, and education sciences as a main index of the reliability and internal consistency (Bollen, 1989, p.215\cite{Bollen1989Structural}; Klaas, 2009\cite{Klaas2009Correcting}). Due to its extensive application in practice, coefficient alpha has generated a great deal of discussion and evaluation on its use, abuse, virtues, limitation and comparison with other methods for test reliability (Schmitt, 1996\cite{Schmitt1996Uses}; Klaas, 2009\cite{Klaas2009Correcting}; Tavakol and Dennick, 2011\cite{Tavakol2011Making}; Zinbarg, Revelle, Yovel, and Li, 2005\cite{Zinbarg2005THEIR}). Meanwhile, development of the methodology for coefficient alpha has been going on over several decades to meet the need in its application. (Woodruff and Feldt, 1986\cite{Woodruff1986Tests}) presented and evaluated several statistical procedures to test the equality of $m$ coefficient alphas from dependent samples. (Feldt and Ankenmann, 1999\cite{Feldt1999Determining}; Bonett, 2002\cite{Bonett2002Sample}) developed methods for constructing confidence intervals and testing for coefficient alpha, as well as methods for determining the sample size required to attain desired power for test of coefficient alpha. (Raykov, West and Traynor, 2015\cite{Raykov2014Evaluation}) proposed a method for point and interval estimation of coefficient alpha for complex sample design based on latent variable modeling. And (Zhang and Yuan, 2016\cite{Zhang2016Robust}) developed a robust procedure to estimate covariance matrix of the responses and coefficient alpha. Their method ease the influence of outlying observations on estimation and can deal with missing data in the responses.

However, none of the aforementioned techniques are purposely developed for or can deal with the situation that internal consistency of the responses varies with the heterogeneity of the subjects. In another word, existing statistical methods only consider summary analysis of coefficient alpha, thus are deficient in assessing the results of the investigation research, where internal consistency of the test score is considered to be associated with some recorded characteristics of the participants. In such studies, for example, (Gilmour et al, 1997\cite{Gilmour1997Measuring})'s Cervical Ectopy Study, (Feinleib et al, 1977\cite{Feinleib1977THE})'s NHLBI Veteran Twin Study, and (Shen et al, 2015\cite{Shen2015Assessment})'s Health Literacy Study in South China, reliability assessment is required to be performed considering the potential variation of the responses' internal consistency from subjects to subjects that are characterized by some covariates.

In this article, we introduce individualized coefficient alpha based on the form of overall coefficient alpha, defined to measure the internal consistency specified by the subject-specific covariates of our interests. And we propose a regression modelling method based on three sets of generalized estimating equations (Liang and Zeger, 1986\cite{Liang1986Longitudinal}; Prentice, 1988\cite{Prentice1988Correlated}) that estimates our defined individualized coefficient alpha and models its association with the covariates. Similar work has been done for the kappa coefficient by (Williamson, Lipsitz and Manatunga, 2000\cite{Williamson2000Modeling}). We discuss asymptotic property of the estimators in our model, on which techniques of interval estimation and hypothesis tests for the regression coefficients and coefficient alpha are based. In addition, we accommodate nonmonotone missingness of the key covariate in our method.

The remainder of the article is organized as follows. Section \ref{sec_not} introduces definition and notation in this article. Section \ref{sec_method} presents formulation of our model and its estimation procedure. Section \ref{cons} demonstrates consistent estimation of the parameters and coefficient alpha in our model. Section \ref{sec_sim} presents and evaluates performance of our proposed method in simulation studies. Section \ref{sec_real} illustrates our methods' application on the Health Literacy Dataset, from (Shen et al, 2015\cite{Shen2015Assessment})'s study in Hunan province of China.

\section{Notation}\label{sec_not}
Suppose that $n$ subjects are scored on $k$ items and let $Y_{ip}$ denote the score of the subject $p$ on the item $i$. $Y_{ip}$ is a either discrete or continuous outcome. Denote that $\mu_{ip}=E[Y_{ip}|{\bf X}_{ip}]$ and variance $\sigma^2_{ip}={\rm Var}[Y_{ip}|{\bf Z}_{ip}]$, where ${\bf X}_{ip}$ and ${\bf Z}_{ip}$ are some covariates of our concerned. Let $S=\sum_{i=1}^kY_i$, and the definition of the population overall coefficient alpha is:
\begin{equation}
\alpha=\left(\frac{k}{k-1}\right)\left(1-\frac{\sum_{i=1}^k{\rm Var}(Y_i)}{{\rm Var}(S)}\right)=\frac{2k\sum_{i\neq j}{\rm Cov}(Y_i,Y_j)}{(k-1){\rm Var}(S)},
\label{equ:1}
\end{equation}
which is equivalent to
\begin{equation}
\frac{(k-1)\alpha}{k-(k-1)\alpha}=\frac{2\sum_{i\neq j}{\rm Cov}(Y_i,Y_j)}{\sum_{i=1}^k{\rm Var}(Y_i)}=\frac{2\sum_{i\neq j}{\rm E}Y_iY_j-{\rm E}Y_i{\rm E}Y_j}{\sum_{i=1}^k{\rm Var}(Y_i)}.
\label{equ:2.1}
\end{equation}
To extend it to be a subject and item specified measurement of the internal consistency, we refer to equation (\ref{equ:2.1}) to define $\alpha_{ijp}$, coefficient alpha of the subject $p$ and the pair of items $(i,j)$, which is introduced to quantify the internal consistency specified by subjects and items. Let
\begin{equation}
\frac{\alpha_{ijp}}{2-\alpha_{ijp}}={\rm E}\left[\frac{2(Y_{ip}Y_{jp}-\mu_{ip}\mu_{jp})}{\sigma_{ip}^2+\sigma_{jp}^2}\Bigg|{\bf W}_{ijp}\right],
\label{equ:4}
\end{equation}
and $\alpha_{ijp}$ is defined by equation (\ref{equ:4}). Here we reduce all of the $k$ items in equation (\ref{equ:2.1}) to the pair $(i,j)$, and naturally extend the variance and covariance of the responses to the expectation of $Y_{ip}Y_{jp}$ conditioned on ${\bf W}_{ijp}$, the covariates of our interests. To allow wide use of our method, we also consider missingness of covariates in the data. For the subject $p$, let $\delta_p$ denote an indicator for the missingness of the covariate that $\delta_p=0$ for having missing values, and $\delta_p=1$, otherwise.

\section{Model}
Assume that $\mu_{ip}$ is associated with covariates ${\bf X}_{ip}$ through link function $g$ and the parameter $\beta$:
\begin{equation}
\mu_{ip}=\mu_{ip}(\beta):=g({\bf X}^T_{ip}\beta).
\label{equ:2}
\end{equation}
Noticing that $\sigma^2_{ip}$ may not be determined by $\mu_{ip}$, for example, when some latent random effects exists to affect $Y_{ip}$, or $Y_{ip}$ is normally distributed marginally, we also parameterize it with function $h$ and parameter $\omega$ because it is not a nuisance parameter when we are modeling coefficient alpha:
\begin{equation}
\sigma^2_{ip}=\sigma^2_{ip}(\omega):=h({\bf Z}^T_{ip}\omega).
\label{equ:3}
\end{equation}
The range of our newly defined $\alpha_{ijp}$ is $(-\infty,1]$ that is the same as $\alpha$. To avoid the restriction of the parameter space, we use the Fisher's z-transformation:
\begin{equation}
\frac{\alpha_{ijp}}{2-\alpha_{ijp}}=\eta_{ijp}(\theta):=\frac{1-{\rm exp}({\bf W}^T_{ijp}\theta)}{1+{\rm exp}({\bf W}^T_{ijp}\theta)},
\label{equ:5}
\end{equation}
where $\theta$ is the regression coefficients, and it implies that $\alpha_{ijp}=1-{\rm exp}({\bf W}^T_{ijp}\theta)$. In practice, we are more concerned about following parameters:
\begin{itemize}
\item $\alpha_{ijp}$, measuring the internal consistency of $p$'s scores on the items $i$ and $j$.
\item Coefficients $\theta$, reflecting the relationship between the value of the individualized coefficient alpha $\alpha_{ijp}$ and the covariate ${\bf W}_{ijp}$.
\end{itemize}
For missing data, we denote that $\pi_p={\rm P}(\delta_p=1|{\bf Q}_p)$, and assume
\begin{equation}
{\rm logit}(\pi_p)={\bf Q}_p^T\gamma,
\label{equ:2.2}
\end{equation}
where ${\bf Q}_p$ is covariates related to the missingness of the data, and $\gamma$ is the coefficients for ${\bf Q}_p$. Our goal of modelling missing data is to obtain unbiased and low variance estimators for the parameters of our concern.
\section{Estimation}\label{sec_method}
Since the joint distribution of the responses $(Y_{1p},Y_{2p},\cdots,Y_{kp})$ is not specified, we propose a method based on (Liang and Zeger, 1986\cite{Liang1986Longitudinal})'s generalized estimating equation (GEE) approach to model the data, and to estimate the parameters $\beta$, $\omega$ and $\theta$. Generalized estimating equation is an extension of general linear model, and is useful to analyze correlated responses when their distribution is not fully specified. Enlightened by (Prentice, 1988\cite{Prentice1988Correlated})'s two-set-GEE method, we propose a three-set-of GEE method to achieve our goal of modeling the individualized coefficient alpha. In order to obtain an unbiased estimators from the GEE, we refer to (Toledano and Gatsonis, 1999\cite{Toledano1999Generalized})'s methods of processing missing data in generalized estimating equations. Assume the missing data mechanism is missing at random conditional on the observed covariate ${\bf Q}_p$, and we reweight each set of GEE with the inverse of observed probability $\pi_p$, which is related to ${\bf Q}_p$ as equation (\ref{equ:2.2}).

Denote that ${\bf Y}_p=(Y_{1p},Y_{2p},\cdots, Y_{kp})^T$ and $\boldsymbol{\mu}_p=(\mu_{1p},\mu_{2p},\cdots,\mu_{kp})^T$. Firstly, we introduce the first set of estimating equations to estimate $\beta$:
\begin{equation}
e_{1}(\beta)=:\sum_{p=1}^n\frac{\delta_p}{\pi_p}{\bf D}^T_p{\bf V}_{1p}^{-1}({\bf Y}_p-\boldsymbol{\mu}_p)=0,
\label{equ:6}
\end{equation}
where ${\bf D}_p=d\boldsymbol{\mu}_p(\beta)/d\beta$, ${\bf V}_{1p}={\bf V}_{1p}(\beta,\nu_1)$ is the working covariance matrix of ${\bf Y}_p$, and $\nu_1$ is the nuisance covariance parameter. According to (Liang and Zeger, 1986\cite{Liang1986Longitudinal}), consistent property of the parameter estimation is guaranteed with no need to correctly estimate $\nu_1$ and $V_{1p}$. Noting from equation (\ref{equ:4}) that estimation of $Y_{ip}$'s variance is necessary for estimating the coefficient alpha, we propose the second set of estimating equation to estimate $\sigma^2_{ip}$. Let ${\bf T}_p=(T_{1p},T_{2p},\cdots, T_{kp})^T$ where $T_{ip}=(Y_{ip}-\mu_{ip})^2$, and ${\bf T}_p$'s conditional expectation be $\boldsymbol{\sigma}^2_p=(\sigma^2_{1p},\sigma^2_{2p},\cdots,\sigma^2_{kp})^T$. We have
\begin{equation}
e_{2}(\beta,\omega)=:\sum_{p=1}^n\frac{\delta_p}{\pi_p}{\bf E}^T_p{\bf V}_{2p}^{-1}({\bf T}_p-\boldsymbol{\sigma}^2_p)=0,
\label{equ:7}
\end{equation}
where ${\bf E}_p=d\boldsymbol{\sigma}^2_p(\omega)/d\omega$, and ${\bf V}_{2p}={\bf V}_{2p}(\beta,\omega,\nu_2)$ is also the working covariance matrix. Then we let
\begin{equation}
U_{ijp}=\frac{2(Y_{ip}Y_{jp}-\mu_{ip}\mu_{jp})}{\sigma_{ip}^2+\sigma_{jp}^2},
\label{equ:8}
\end{equation}
${\bf U}_p=(U_{12p},U_{13p},\cdots,U_{(k-1)kp})^T$, and $\boldsymbol{\eta}_p=(\eta_{12p},\eta_{13p},\cdots,\eta_{(k-1)kp})^T$. The third set of estimating equations is proposed to estimate $\theta$ and the individualized coefficient alpha:
\begin{equation}
e_{3}(\beta,\omega,\theta)=:\sum_{p=1}^n\frac{\delta_p}{\pi_p}{\bf F}^T_p{\bf V}_{3p}^{-1}({\bf U}_p-\boldsymbol{\eta}_p)=0,
\label{equ:9}
\end{equation}
where ${\bf F}_p=d\boldsymbol{\eta}_p(\theta)/d\theta$ and the working covariance ${\bf V}_{3p}={\bf V}_{3p}(\beta,\omega,\theta,\nu_3)$.

To solve the three sets of GEE, we firstly implement logistic regression to estimate $\gamma$. Denote the estimator as $\widehat{\gamma}$ and $\widehat{\pi}_p={\rm logit}^{-1}({\bf Q}_p\widehat{\gamma})$. Then we implement Gauss-Newton algorithm to compute $(\widehat{\beta},\widehat{\omega},\widehat{\theta})$, estimators for $(\beta,\omega,\theta)$. The three set of GEE are solved successively. In the $m$th iteration of solving each set of equations, we update the parameters by
\begin{equation}
\beta^{(m+1)}=\beta^{(m)}-\left(\sum_{p=1}^n\frac{\delta_p}{\widehat{\pi}_p}{\bf D}^T_p{\bf V}_{1p}^{-1}{\bf D}_p\right)^{-1}\left[\sum_{p=1}^n\frac{\delta_p}{\widehat{\pi}_p}{\bf D}^T_p{\bf V}_{1p}^{-1}({\bf Y}_p-\boldsymbol{\mu}_p)\right]\Bigg|_{\beta=\beta^{(m)},\nu_1=\nu_1^{(m)}},
\label{equ:10}
\end{equation}
\begin{equation}
~~~~~~~~~\omega^{(m+1)}=\omega^{(m)}-\left(\sum_{p=1}^n\frac{\delta_p}{\widehat{\pi}_p}{\bf E}^T_p{\bf V}_{2p}^{-1}{\bf E}_p\right)^{-1}\left[\sum_{p=1}^n\frac{\delta_p}{\widehat{\pi}_p}{\bf E}^T_p{\bf V}_{2p}^{-1}({\bf T}_p-\boldsymbol{\sigma}^2_p)\right]\Bigg|_{(\beta,\omega)=(\widehat{\beta},\omega^{(m)}),\nu_2=\nu_2^{(m)}}
\label{equ:11}
\end{equation}
and
\begin{equation}
~~~~~~~~~~~~~~\theta^{(m+1)}=\theta^{(m)}-\left(\sum_{p=1}^n\frac{\delta_p}{\widehat{\pi}_p}{\bf F}^T_p{\bf V}_{3p}^{-1}{\bf F}_p\right)^{-1}\left[\sum_{p=1}^n\frac{\delta_p}{\widehat{\pi}_p}{\bf F}^T_p{\bf V}_{3p}^{-1}({\bf U}_p-\boldsymbol{\eta}_p)\right]\Bigg|_{(\beta,\omega,\theta)=(\widehat{\beta},\widehat{\omega},\theta^{(m)}),\nu_3=\nu_3^{(m)}},
\label{equ:12}
\end{equation}
while the nuisance parameters $\nu_1$, $\nu_2$ and $\nu_3$ are updated by the method of moments in each iteration. In this way, $\beta$ is estimated by iteratively implement (\ref{equ:10}) and updating $\nu_1$. Then we obtain $\widehat{\omega}$ in similar procedure using $\widehat{\beta}$. And $\theta$ is estimated based on $\widehat{\beta}$ and $\widehat{\omega}$. Since $e_{1}(\beta)$ is not a function of $\omega$ and $\theta$, and $e_{2}(\beta,\omega)$ is not a function of $\theta$, there is no need to go back and forth between these three set of estimating equations. This point is similar to the methods proposed by (Prentice, 1988\cite{Prentice1988Correlated}) and (Williamson, Lipsitz, and Manatunga, 2000\cite{Williamson2000Modeling}).

With $\widehat{\theta}$ estimated by solving equation (\ref{equ:7}), (\ref{equ:8}) and (\ref{equ:9}), $\alpha_{ijp}$ is estimated by $\widehat{\alpha}_{ijp}=1-{\rm exp}({\bf W}^T_{ijp}\widehat{\theta})$. The asymptotic normality of $\widehat{\beta}$, $\widehat{\omega}$ and $\widehat{\theta}$ is given by theorem $\ref{the:1}$ in section \ref{cons}. Then, by theta method, we can also prove the asymptotic normality of $\widehat{\alpha}_{ijp}$, and estimate their asymptotic variance.

\section{Asymptotic Property of Estimators}\label{cons}
In this section, we present and prove the result that the joint asymptotic distribution of $\sqrt{n}(\widehat{\beta}-\beta)$, $\sqrt{n}(\widehat{\omega}-\omega)$ and $\sqrt{n}(\widehat{\theta}-\theta)$ is multivariate Gaussian with mean zero. Let
\begin{equation}
{\bf G}_p=\left(
  \begin{array}{ccc}
    {\bf D}_p&{\bf 0}&{\bf 0}\\
    {\bf 0}&{\bf E}_p&{\bf 0}\\
    {\bf 0}&{\bf 0}&{\bf F}_p\\
  \end{array}
\right),~~~~
{\bf V}_p=\left(
  \begin{array}{ccc}
    {\bf V}_{1p}&{\bf 0}&{\bf 0}\\
    {\bf 0}&{\bf V}_{2p}&{\bf 0}\\
    {\bf 0}&{\bf 0}&{\bf V}_{3p}\\
  \end{array}
\right),~~~~
{\bf f}_p=\left(
  \begin{array}{ccc}
    {\bf Y}_p-\boldsymbol{\mu}_p\\
    {\bf T}_p-\boldsymbol{\sigma}^2_p\\
    {\bf U}_p-\boldsymbol{\eta}_p\\
  \end{array}
\right),
\label{equ:13}
\end{equation}
and the three sets of generalized estimating equations can be formulated jointly by
\begin{equation}
\sum_{p=1}^n\frac{\delta_p}{\pi_p}{\bf G}^T_p{\bf V}_{p}^{-1}{\bf f}_p=0.
\label{equ:14}
\end{equation}
The following theorem gives the large sample property of the regression coefficients:
\begin{theorem}
Denote that
\begin{equation}
{\bf H}_p=\left(
  \begin{array}{ccc}
    d\boldsymbol{\mu}_p/d\beta&{\bf 0}&{\bf 0}\\
    -d{\bf T}_p/d\beta&d\boldsymbol{\sigma}^2_p/d\omega&{\bf 0}\\
    -d{\bf U}_p/d\beta&-d{\bf U}_p/d\omega&d\boldsymbol{\eta}_p/d\theta\\
  \end{array}
\right).
\label{equ:16}
\end{equation}
Assume that the missing data mechanism is missing at random conditional on ${\bf Q}_p$, and $\pi_p$ is bounded in probability away from zero. Under mild regularity conditions, and given that the estimator for $\nu_l$ is $\sqrt{n}$-consistent given $\beta$ when $l=1$, given $(\beta,\omega)$ when $l=2$, and given $(\beta,\omega,\theta)$ when $l=3$, the estimator of equation (\ref{equ:14}) is $\sqrt{n}$-consistent to $(\beta^T,\omega^T,\theta^T)^T$ that $\sqrt{n}((\widehat{\beta}-\beta)^T,(\widehat{\omega}-\omega)^T,(\widehat{\theta}-\theta)^T)^T$ is asymptotically multivariate Gaussian with the mean to be zero, and covariance matrix
\begin{equation}
\begin{split}
{\bf \Psi}=\underset{n\rightarrow\infty}{{\rm lim}}{\bf \Gamma}_n^{-1}({\bf \Sigma}_n-{\bf \Upsilon}_n{\bf \Omega}_n{\bf \Upsilon}_n^T){\bf \Gamma}_n^{-1},
\end{split}
\label{equ:15}
\end{equation}
where
\begin{equation}
\begin{split}
{\bf \Gamma}_n&=\frac{1}{n}\sum_{p=1}^n\frac{\delta_p}{\pi_p}{\bf G}^T_p{\bf V}_{p}^{-1}{\bf H}_p,\\
{\bf \Sigma}_n&=\frac{1}{n}\sum_{p=1}^n(\frac{\delta_p}{\pi_p})^2{\bf G}^T_p{\bf V}_{p}^{-1}{\bf f}_p{\bf f}_p^T{\bf V}_{p}^{-1}{\bf G}_p,\\
{\bf \Upsilon}_n&=\frac{1}{n}\sum_{p=1}^n(\frac{\delta_p}{\pi_p})^2{\bf G}^T_p{\bf V}_{p}^{-1}{\bf f}_p\frac{\partial\pi_p^T}{\partial\gamma},\\
{\bf \Omega}&_n=\frac{1}{n}\sum_{p=1}^n\frac{1}{\pi_p(1-\pi_p)}\frac{\partial\pi_p}{\partial\gamma}\frac{\partial\pi_p^T}{\partial\gamma}.
\end{split}
\label{equ:4.1}
\end{equation}
\label{the:1}
\end{theorem}
Theorem \ref{the:1} indicates that the asymptotic variance of estimators in our model can be consistently estimated with the estimators, the covariates and the responses. Referring to (Pierce, 1982\cite{Pierce1982The}; Liang and Zeger, 1986\cite{Liang1986Longitudinal}; Toledano and Gatsonis, 1999\cite{Toledano1999Generalized}), we prove it as follow:
\begin{proof}
Let $\nu=(\nu_1^T,\nu_2^T,\nu_3^T)^T$, $\zeta=(\beta^T,\omega^T,\theta^T)^T$, ${\bf S}_n(\zeta,\nu,\gamma)=1/n\sum_{p=1}^n(\delta_p/\pi_p){\bf G}^T_p{\bf V}_{p}^{-1}{\bf f}_p$, and $m$ be the total number of the covariates included in the equation (\ref{equ:14}). Denote the $\sqrt{n}$-consistent estimator of $\nu$ given $\zeta$ as $\widehat{\nu}:=\widehat{\nu}(\zeta)$. Under some regularity conditions,
\begin{equation}
\sqrt{n}(\widehat{\zeta}-\zeta)=-\left[\frac{\partial{\bf S}_n(\zeta,\widehat{\nu}(\zeta),\widehat{\gamma})}{\partial\zeta}+\frac{\partial{\bf S}_n(\zeta,\widehat{\nu}(\zeta),\widehat{\gamma})}{\partial\nu}\frac{d\widehat{\nu}(\zeta)}{d\zeta}\right]^{-1}\sqrt{n}{\bf S}_n(\zeta,\widehat{\nu},\widehat{\gamma})
\label{equ:5.0}
\end{equation}
Following a first-order expansion, we also have
\begin{equation}
\begin{split}
\sqrt{n}{\bf S}_n(\zeta,\widehat{\nu},\widehat{\gamma})&=\sqrt{n}{\bf S}_n(\zeta,\widehat{\nu},\gamma)+{\bf B}_1\sqrt{n}(\widehat{\gamma}-\gamma)+o_m(1)\\
&=\sqrt{n}{\bf S}_n(\zeta,\nu,\gamma)+{\bf B}_1\sqrt{n}(\widehat{\gamma}-\gamma)+{\bf B}_2\sqrt{n}(\widehat{\nu}-\nu)+o_m(1),
\end{split}
\label{equ:5.11}
\end{equation}
where ${\bf B}_1=\partial {\bf S}_n(\zeta,\widehat{\nu},\gamma)/\partial\gamma$, and ${\bf B}_2=\partial {\bf S}_n(\zeta,\nu,\gamma)/\partial\nu$. Since the estimating equations are unbiased (the MAR condition), and $\pi_p$ is bounded in probability away from zero, we have
\begin{equation}
{\rm E}\left[\frac{\partial}{\partial\nu}\left(\frac{\delta_p}{\pi_p}{\bf G}^T_p{\bf V}_{p}^{-1}{\bf f}_p\right)\right]={\rm E}\left[\left(\frac{\delta_p}{\pi_p}{\bf G}^T_p\frac{\partial{\bf V}_{p}^{-1}}{\partial\nu}{\bf f}_p\right)\right]=0,
\label{equ:5.2}
\end{equation}
which indicates that ${\bf B}_2=o_m(1)$. With the assumption that $\widehat{\nu}$ is $\sqrt{n}$-consistent to $\nu$ given $\zeta$, we have
\begin{equation}
{\bf B}_2\sqrt{n}(\widehat{\nu}-\nu)=o_m(1)\cdot O_m(1)=o_m(1).
\label{equ:5.3}
\end{equation}
Noting that ${\rm lim}_{n\rightarrow\infty}{\bf \Sigma}_n$ and ${\rm lim}_{n\rightarrow\infty}{\bf \Omega}_n$ are the asymptotic covariance matrix of $\sqrt{n}{\bf S}_n(\zeta,\nu,\gamma)$ and $\sqrt{n}(\widehat{\gamma}-\gamma)$ respectively, then by equation (\ref{equ:5.11}), results of (Pierce, 1982\cite{Pierce1982The}) give that
\begin{equation}
\sqrt{n}{\bf S}_n(\zeta,\widehat{\nu},\widehat{\gamma})\xrightarrow{f}L\sim{\rm N}\left({\bf 0},\underset{n\rightarrow\infty}{{\rm lim}}({\bf \Sigma}_n-{\bf B}_1{\bf \Omega}_n{\bf B}_1^T)\right),
\label{equ:5.4}
\end{equation}
where ${\rm lim}_{n\rightarrow\infty}{\bf B}_1={\rm lim}_{n\rightarrow\infty}{\bf \Upsilon}_n$, according to our assumptions. Meanwhile, it is not difficult to show that $d\widehat{\nu}(\zeta)/d\zeta$ is bounded, that ${\rm lim}_{n\rightarrow\infty}\partial{\bf S}_n(\zeta,\widehat{\nu}(\zeta),\widehat{\gamma})/{\partial\zeta}={\rm lim}_{n\rightarrow\infty}{\bf \Gamma}_n$, and that ${\rm lim}_{n\rightarrow\infty}\partial{\bf S}_n(\zeta,\widehat{\nu}(\zeta),\widehat{\gamma})/{\partial\nu}={\rm lim}_{n\rightarrow\infty}{\bf B}_2={\bf 0}$. Substitute these and the equation (\ref{equ:5.4}) into the equation (\ref{equ:5.0}), and we complete the proof.
\end{proof}

\section{Simulation Study}\label{sec_sim}
\subsection{Simulation Settings}\label{set}
To assess the performance of our method in estimating parameters, evaluating the coefficient alpha and detecting significance, we conduct series of simulation studies. To mimic Health Literacy dataset (see section \ref{sec_real}), we set the number of items to be 3 and sample size to be 2500, 3000 and 3500, and generate 500 datasets for each set of sample size. Number and distribution of the covariates are also made close to the real data, too. Specifically, we let the ${\bf Y}_p=(Y_{1p},Y_{2p},Y_{3p})^T$ be multivariate Gaussian and
\begin{equation}
{\rm E}[Y_{ip}|{\bf X}_{ip}]=\beta_{0i}+x_{1p}\beta_1+x_{2ip}\beta_2+x_{3ip}\beta_3,
\label{equ:17}
\end{equation}
where $\beta_{01}=-0.6$, $\beta_{02}=0.4$, $\beta_{03}=0.3$, $\beta_1=0.25$, $\beta_2=0$, $\beta_3=1.0$, and covariates ${\bf X}_{ip}=(x_{1p},x_{2ip},x_{3ip})^T$ generated by $x_{1p}, x_{2ip}\sim {\rm Unif}(-1, 1)$, $x_{3ip}\sim {\rm N}(0, 1)$. Set ${\rm var}(Y_{ip}|{\bf X}_{ip})=1$ and
\begin{equation}
{\rm cov}(Y_{ip},Y_{jp}|{\bf W}_{ijp},{\bf X}_{ip},{\bf X}_{jp})=[1-{\rm exp}({\bf W}_{ijp}^T\theta)]/[1+{\rm exp}({\bf W}_{ijp}^T\theta)],
\label{equ:18}
\end{equation}
where
\begin{equation}
{\bf W}_{ijp}^T\theta=\theta_0+w_{1p}\theta_1+w_{2p}\theta_2+w_{3p}\theta_3+w_{4ijp}\theta_4+w_{5ijp}\theta_5,
\label{equ:19}
\end{equation}
and $\theta=(-0.6, -0.4, 0.05, 0.05, -0.2, 0)^T$. We generate ${\bf W}_{ijp}=(w_{1p},w_{2p},w_{3p},w_{4ijp},w_{5ijp})^T$ by $w_{1p}\sim{\rm Bern}(0.5)$, $w_{2p}, w_{4ijp}\sim {\rm Unif}(0,1)$, $w_{3p}, w_{5ijp}\sim {\rm N}(0,1)$. In this way, the coefficient alpha specified by subject $p$ and pair of item $(i,j)$ is given by $\alpha_{ijp}=1-{\rm exp}({\bf W}_{ijp}^T\theta)$ according to our definition in section \ref{sec_method}. In addition, we generate random missingness of the covariates of subject $p$ with the probability of verification $\pi_p$ satisfying
\begin{equation}
{\rm logit}(\pi_p)=\gamma_0+q_{1p}\gamma_1+q_{2p}\gamma_2,
\label{equ:5.1}
\end{equation}
where $\gamma=(2, 0.5, -0.6)^T$, and both $q_{1p}$ and $q_{2p}$ are generated from ${\rm N}(0,1)$. In this way, the rate of missingness of the key covariate is around 0.8, which is close to the property of Health Literacy dataset, our motivating dataset.

\subsection{Parameter Estimation}
Based on our method's results on the 500 datasets for each set of sample size, we estimate mean value and root mean squared error (RMSE) of the estimators for the regression coefficient $\theta$ in the GEEs. As one of our main concerns, estimators $\widehat{\theta}$ of different settings for sample size are evaluated in Table \ref{tab:1}.
\begin{table}[htbp]
\center
\begin{tabular}{ccccccc}
\hline
Sample size & \multicolumn{2}{c}{2500} & \multicolumn{2}{c}{3000} & \multicolumn{2}{c}{3500} \\
\hline
      & Mean  & RMSE   & Mean  & RMSE   & Mean  & RMSE \\
\hline
$\theta_0(=-0.6)$ & -0.593 & 0.13 & -0.602 & 0.123 & -0.600 & 0.115 \\
$\theta_1(=-0.4)$ & -0.405 & 0.127 & -0.395 & 0.105 & -0.405 & 0.102 \\
$\theta_2(=0.05)$ & 0.038 & 0.178 & 0.054 & 0.176 & 0.059 & 0.160 \\
$\theta_3(=0.05)$ & 0.046 & 0.056 & 0.050 & 0.051 & 0.046 & 0.047 \\
$\theta_4(=-0.2)$ & -0.201 & 0.141 & -0.204 & 0.129 & -0.201 & 0.121 \\
$\theta_5(=0)$ & 0.001 & 0.041 & 0.002 & 0.037 & -0.001 & 0.034 \\
\hline
\end{tabular}%
\caption{\label{tab:1} Mean and RMSE of estimators for the coefficient in the third set of GEEs of our method with the sample size set to be 2500, 3000 and 3500.}
\end{table}
The simulation results in Table \ref{tab:1} demonstrate good performance of our method on estimating the regression coefficients in the third set of GEEs. As the sample size increases from 2500 to 3000, and from 3000 to 3500, mean values of the estimators become closer to the real value of the coefficient, and RMSE of them decrease strictly. Meanwhile, bias and total errors are within an acceptable scale. In summary, our method has good performance in parameter estimation, when the sample size, verification probability and settings of the covariates are similar to Health Literacy dataset. As the sample size increases by a moderate margin, there is an apparent tendency for the estimators being close to the true values. This indicates their good unbiasedness and consistent performance, though distribution of the responses is not specified in our model.

\subsection{Significance Detection}
Two types of hypothesis testing are in our main consideration in our method. One is whether the values of the regression coefficients in the third set of GEEs are significantly different from 0, which reflects the relationship between our defined individualized coefficient alpha and the covariates. Another is to decide whether the coefficient alpha ranges in an acceptable scale, often chosen as $(0.7,0.9)$, as suggested by (Tavakol and Dennich, 2011\cite{Tavakol2011Making}) and (Streiner, 2003\cite{Streiner2003Starting}). According to section \ref{sec_method} and \ref{cons}, these two types of hypothesis testing can be performed based on the estimators' asymptotic normality.

In the simulation study, we evaluate our method's performance on the significant detection under different set of sample size by estimating their type I error rate, and estimate the power when the true values of the parameters changes, or the true value of the coefficient alpha varies with the covariates in our model. Under the settings described in section \ref{set}, we fix the significant level to be 0.05, and estimate the type I error rate of the hypothesis testing $\theta_5=0$ in our 500 groups of simulation with different sample sizes. We also estimate the power of the testing, when the true values of the parameters are respectively $-0.6$, $-0.4$ and $-0.2$. The results are presented in Table \ref{tab:5}, which demonstrates that the type I error rates are well controlled, and get closer to the significant level with the sample size's increasing. The power grows fast with the parameters' absolute values increases, too. These indicates good approximation of the estimators' asymptotic normality estimated by our method to the one with infinite sample size.
\begin{table}[htbp]
\center
\begin{tabular}{cccc}
\hline
\multicolumn{1}{c}{Sample size} & 2500  & 3000  & 3500 \\
\hline
$\theta_0(=-0.6)$  & 99.7\% & 99.8\% & 100\% \\
$\theta_1(=-0.4)$ & 93.6\% & 96.2\% & 98\% \\
$\theta_4(=-0.2)$ & 30.4\% & 34.2\% & 40.6\% \\
$\theta_5(=0)$ & 4.4\% & 5.2\% & 5.0\% \\
\hline
\end{tabular}%
\caption{\label{tab:5} Type one error of the hypothesis testing $\theta_5=0$, and power of the hypothesis testing $\theta_0=0$, $\theta_1=0$, and $\theta_4=0$ with the sample size set to be 2500, 3000 and 3500.}
\end{table}

\begin{figure}[ht]
\centering
    \begin{minipage}[h]{1.9in}
    \centering
    \includegraphics[width=1.9in]{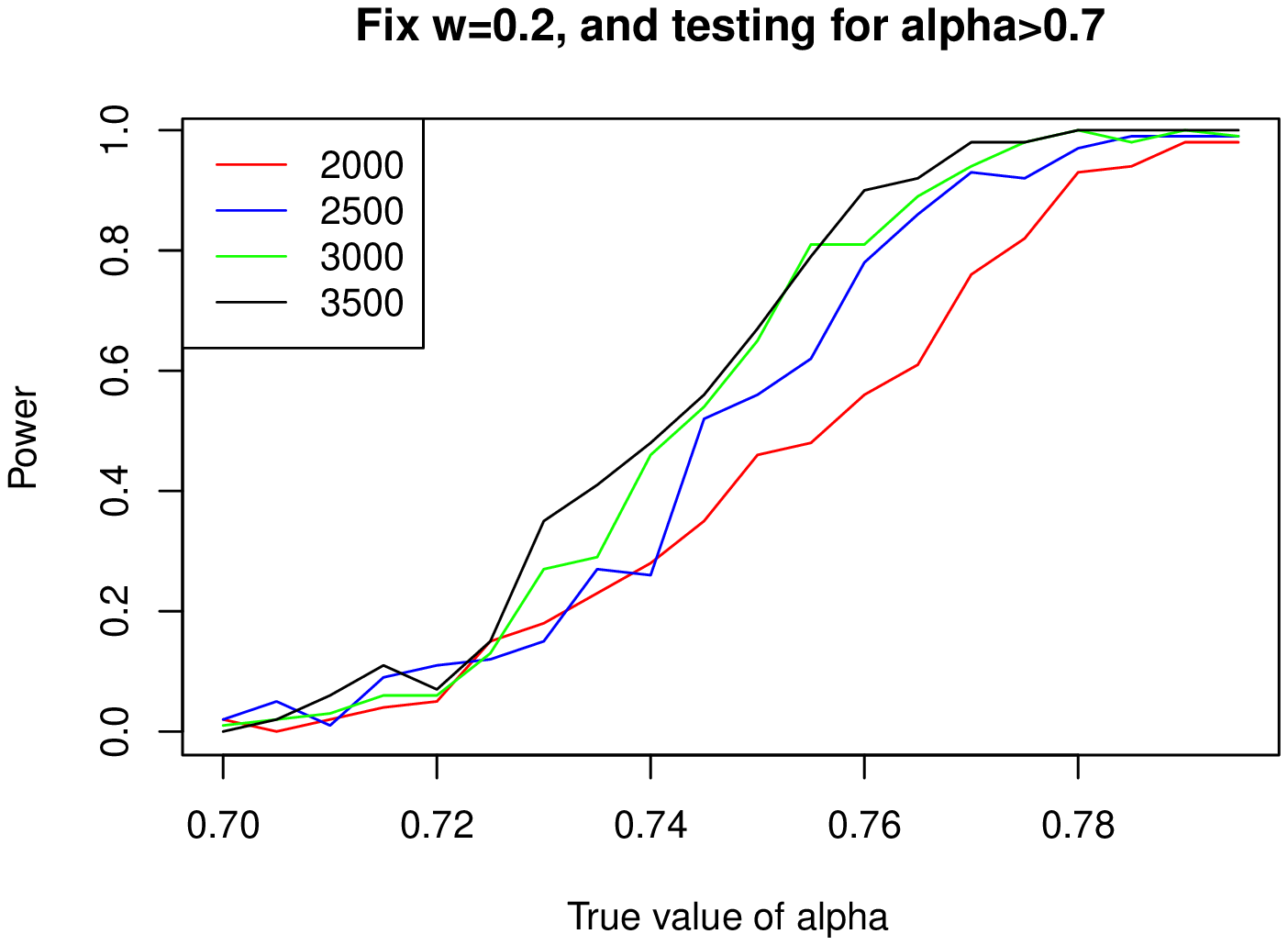}
    \end{minipage}
    \begin{minipage}[h]{1.9in}
    \centering
    \includegraphics[width=1.9in]{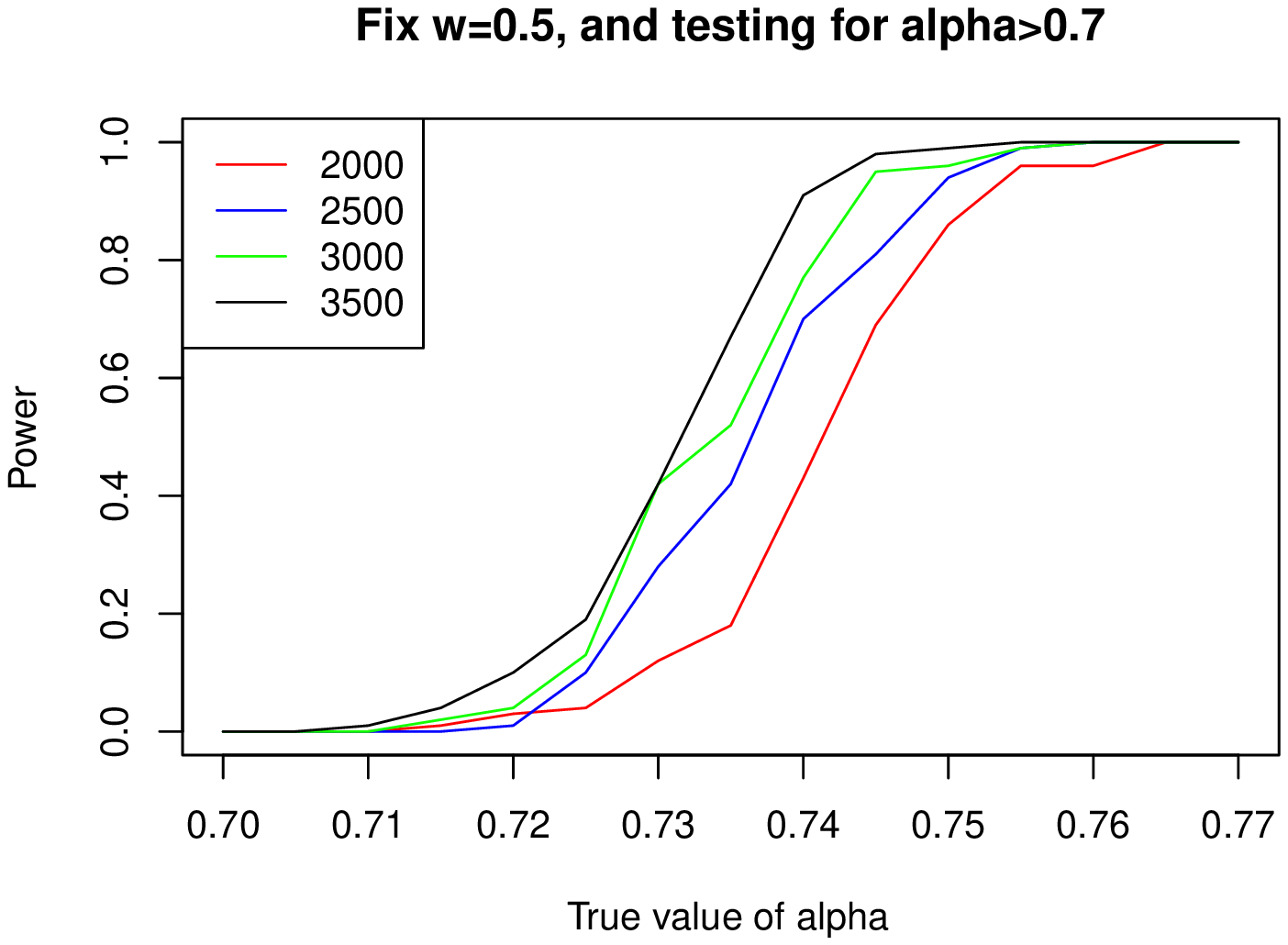}
    \end{minipage}
    \begin{minipage}[h]{1.9in}
    \centering
    \includegraphics[width=1.9in]{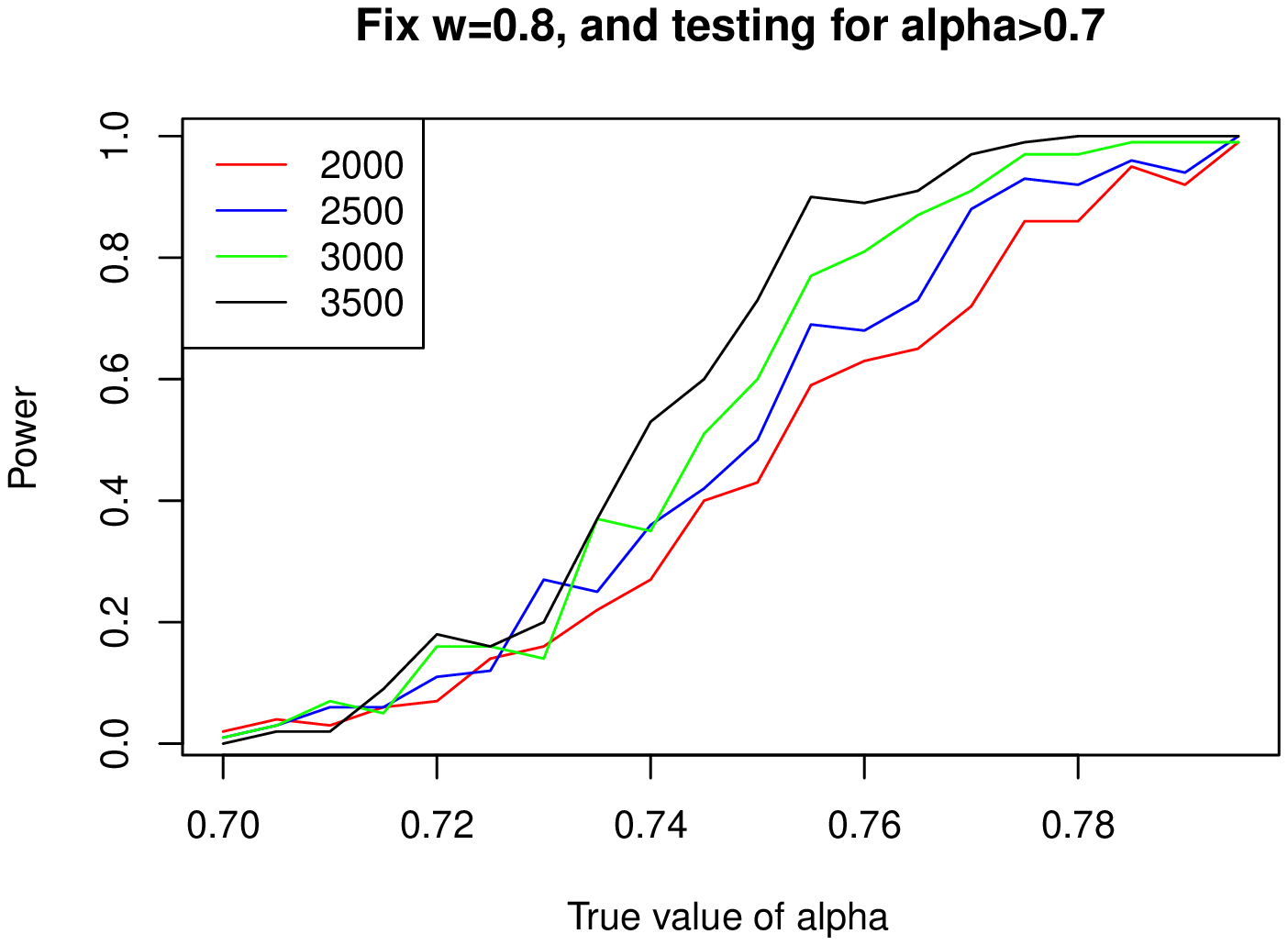}
    \end{minipage}
\caption{\label{fig:1} Set $\widetilde{w}_{1p}=0.2$, $\widetilde{w}_{1p}=0.5$ and $\widetilde{w}_{1p}=0.8$ respectively, plots show the relationship between power of the hypothesis testing $H_0:\widetilde{\alpha}_{ijp}<0.7$ vs $H_1:\widetilde{\alpha}_{ijp}\geq0.7$ and true value of the coefficient alpha $\widetilde{\alpha}_{ijp}$, with the sample size set to be 2000, 2500, 3000 and 3500 respectively.}
\end{figure}

\begin{figure}[ht]
\centering
    \begin{minipage}[h]{1.9in}
    \centering
    \includegraphics[width=1.9in]{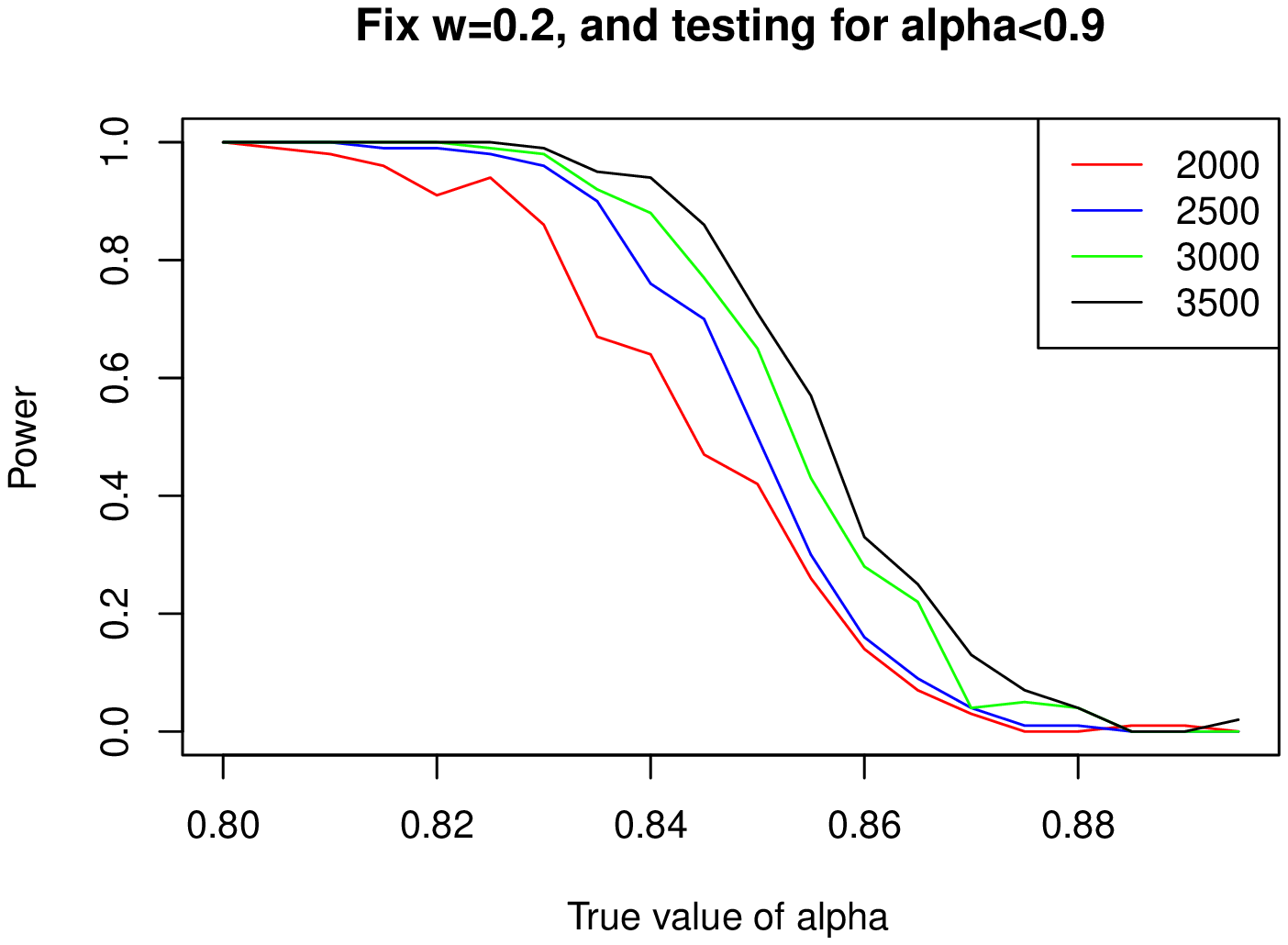}
    \end{minipage}
    \begin{minipage}[h]{1.9in}
    \centering
    \includegraphics[width=1.9in]{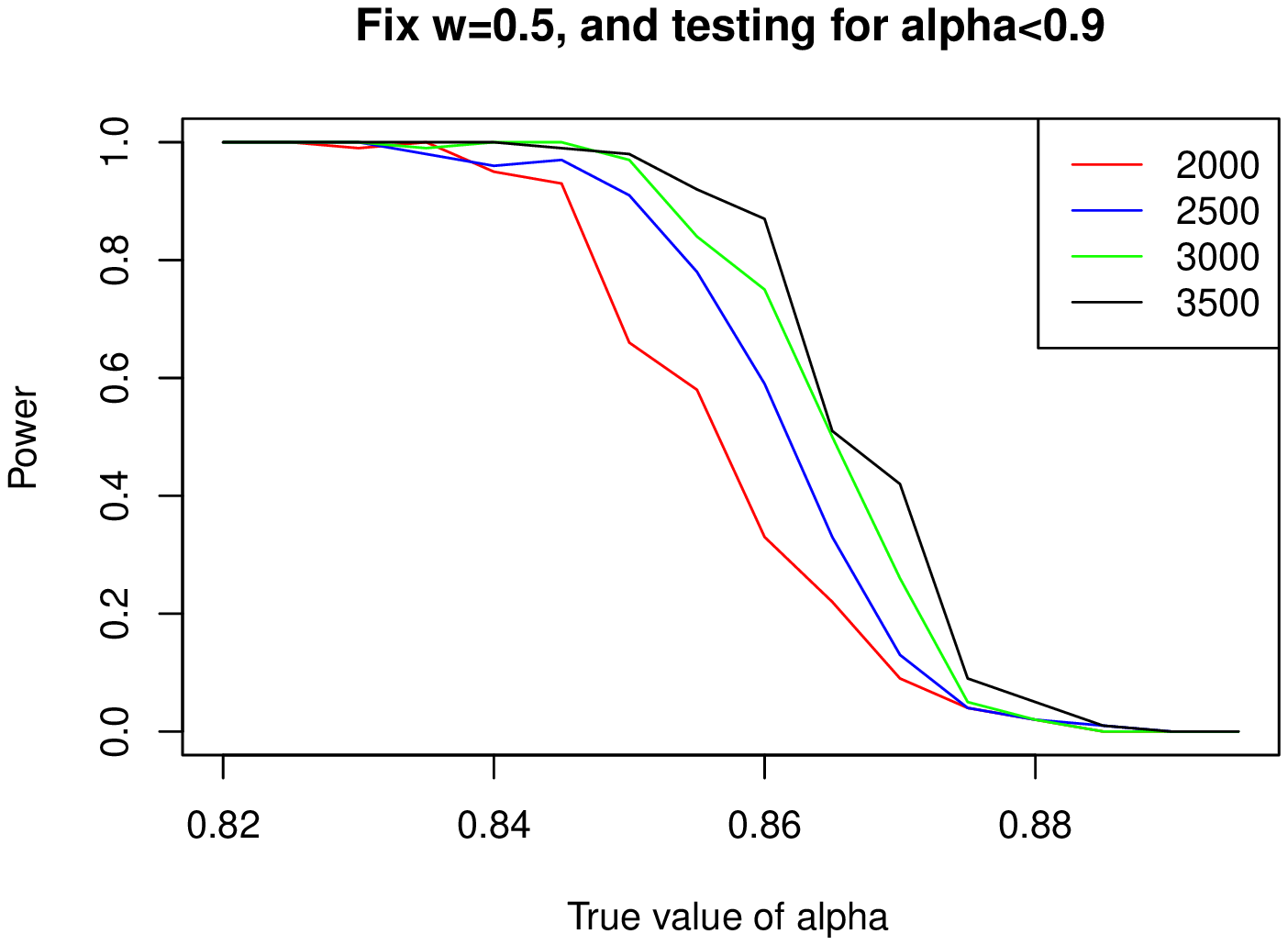}
    \end{minipage}
    \begin{minipage}[h]{1.9in}
    \centering
    \includegraphics[width=1.9in]{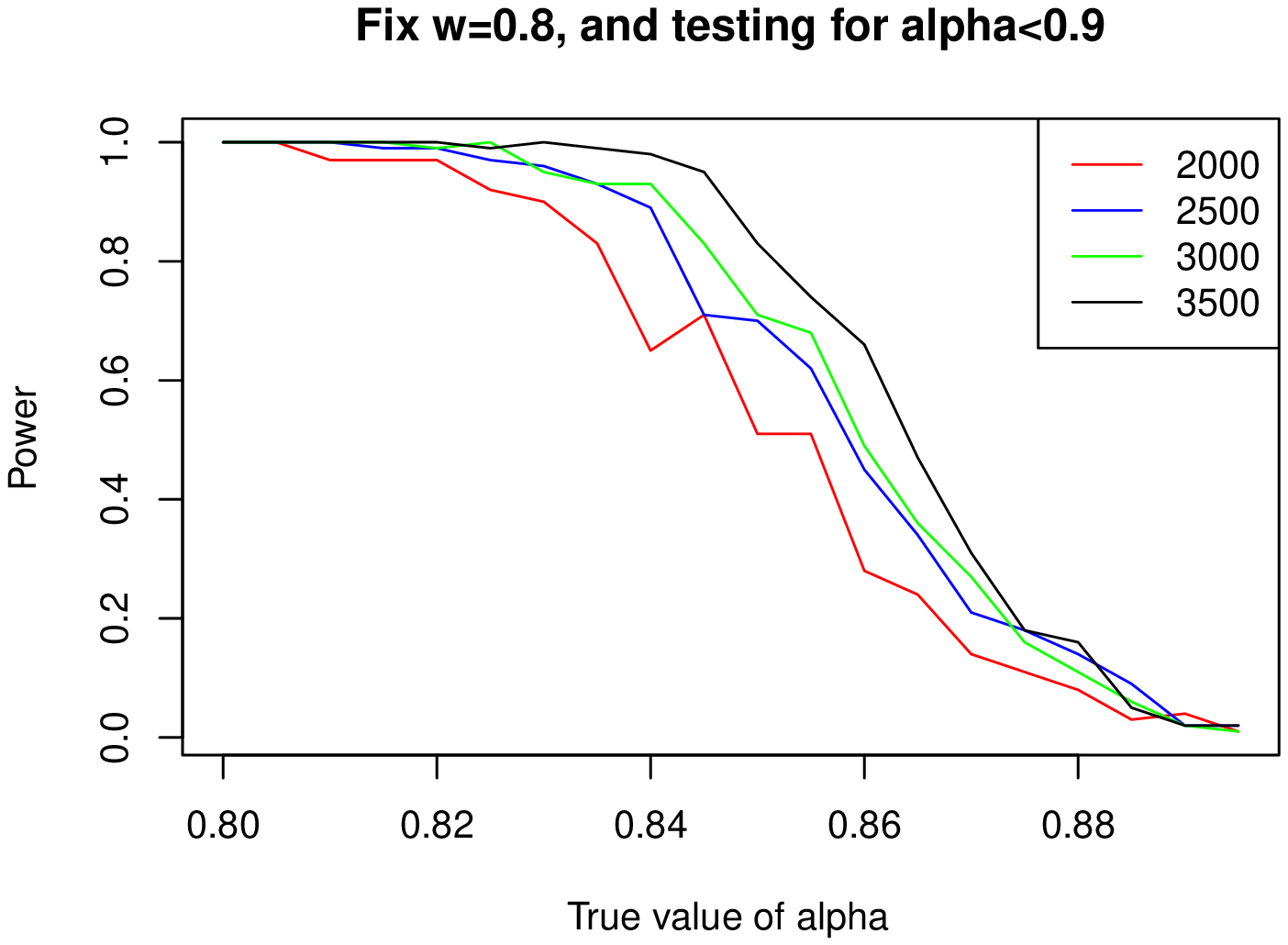}
    \end{minipage}
\caption{\label{fig:2} Set $\widetilde{w}_{1p}=0.2$, $\widetilde{w}_{1p}=0.5$ and $\widetilde{w}_{1p}=0.8$ respectively, plots show the relationship between power of the hypothesis testing $H_0:\widetilde{\alpha}_{ijp}>0.9$ vs $H_1:\widetilde{\alpha}_{ijp}\leq0.9$ and true value of the coefficient alpha $\widetilde{\alpha}_{ijp}$, with the sample size set to be 2000, 2500, 3000 and 3500 respectively.}
\end{figure}

To evaluate our method's power of the hypothesis tests for the range of the coefficient alpha, we set the sample size to be 2000, 2500, 3000 and 3500 respectively, and let $\alpha_{ijp}=1-{\rm exp}(\theta_0+\theta_1w_{1p})$. Covariate $w_{1p}$ is generated from ${\rm Unif}(0,1)$. Fix $\widetilde{w}_{1p}$ to be $0.2$, $0.5$ and $0.8$ respectively, and let $\theta_0={\rm log}(0.3)$. While $\theta_1$ are set properly for each setting of simulation, to make $\widetilde{\alpha}_{ijp}=1-{\rm exp}(\theta_0+\theta_1\widetilde{w}_{1p})$ range from $0.7$ to $0.9$. For each set of sample size and value of $\widetilde{\alpha}_{ijp}$, we estimate our method's powers for the hypothesis testing $H_0:\widetilde{\alpha}_{ijp}<0.7$ vs $H_1:\widetilde{\alpha}_{ijp}\geq0.7$, and the hypothesis testing $H_0:\widetilde{\alpha}_{ijp}>0.9$ vs $H_1:\widetilde{\alpha}_{ijp}\leq0.9$, by performing our methods on 500 simulation datasets. The resulted relationship between the power and true value of $\widetilde{\alpha}_{ijp}$ under different settings are presented in Figure \ref{fig:1} (for $H_0:\widetilde{\alpha}_{ijp}<0.7$ vs $H_1:\widetilde{\alpha}_{ijp}\geq0.7$) and Figure \ref{fig:2} (for $H_0:\widetilde{\alpha}_{ijp}>0.9$ vs $H_1:\widetilde{\alpha}_{ijp}\leq0.9$).

The hypothesis testing for the range of our introduced coefficient alpha is simulated to obtain power of our method for different true values of the coefficient alpha, under two null hypothesis, different sample sizes, and different values of the covariate $\widetilde{w}_{1p}$. In summary, the power increases noticeably with the sample size's growing, and reaches 1 when there is enough difference (says 0.1) between the true value of coefficient alpha with 0.7 or 0.9. These indicate that our estimated asymptotic variance of parameters is within an acceptable scale, and will show a distinct decrease as the sample size varies from 2000 to 2500, 3000 and 3500, which is comparable with the Health Literacy dataset. Also, we can see the fastest rate of the power's being close to 1 when $\widetilde{w}_{1p}=0.5$, which is actually the average value of ${w}_{1p}$. This suggest our method give more reliable results on subjects that are more common in the sample.

\section{Health Literacy Data Analysis}\label{sec_real}
A national health literacy study was conducted by (Shen et al, 2015\cite{Shen2015Assessment}) via a scale-based investigation. Population-based sample of 3731 participants in Hunan Province was included in the study, investigated, and evaluated by the scale on their health literacy. In this section, we apply our proposed method on the resulted Health Literacy Dataset, as an example to illustrate its application in practice. Health Literacy dataset includes three dimensions of health literacy scores, \textsl{knowledge and attitude}, \textsl{behavior and lifestyle}, and \textsl{skills}. Each participant's literacy on each dimension is evaluated via Chinese Resident Health Literacy Scale developed by the investigator. Meanwhile, age ($age$), gender ($gender$), education ($edu$), number of family members ($member$), and income ($income$) of the subjects were also investigated, in the interest of their association with the internal consistency of the three health literacy dimensions (items) in this study. After deletion of those with missing values in more than one variables, totally 3375 subjects are remained, among which 382 are confronted with a missing value for the covariate $income$. With the missingness of $income$ considered, our method is used to model the internal consistency of the health literacy scores on the three dimensions with the investigated covariates. Point and interval estimations of the coefficients in the third set of GEE and our defined coefficient alpha, and significant detection on the relationship between the coefficient alpha and covariates of our interested are presented in this section.

\subsection{Process Missing Data}
For subject $p$, we introduce $\delta_p$ to denote the missingness of its $income$ that $\delta_p=0$ for having a missing value for $income$ and $\delta_p=1$, otherwise. Logistic regression on $\delta_p$ using the other covariates turns out that variables $gender$ and $edu$ are significantly related to the missingness of $income$, while other covariates show no significant association with the verification probability. Denote that ${\rm P}(\delta_p=1|gender,edu)=\pi_p$ and $\widehat{\pi}_p$ be the estimator for $\pi_p$ in the logistic regression, then we have
\begin{equation}
{\rm logit}(\widehat{\pi}_p)=3.26-0.27gender-0.26edu,
\label{equ:20}
\end{equation}
where the p-value of $gender$ and $edu$'s coefficients are 0.01 and less than 0.001 respectively, indicating their significance in the logistic model. Since personal information investigated in the study seems to typically reflect the reasons for missingness in the covariate $income$, the logistic model described above is plausible. And it is reasonable to assume that subject $p$'s $income$ is missing at random conditional on its $gender$ and $edu$, which ensures the use of our method.

\subsection{Results}
With missing data processed, our method is implemented on the Health Literacy dataset to analyze the covariates' association with internal consistency of different items, measured by our defined subject and item specific coefficient alpha. For the first, the second, and the third set of GEEs, we respectively assume that
\begin{equation}
\mu_{ip}=\beta_{0i}+\beta_1age+\beta_2gender+\beta_3edu+\beta_4member+\beta_5income;
\label{equ:22}
\end{equation}
\begin{equation}
\sigma_{ip}^2=\sigma_i^2;
\label{equ:23}
\end{equation}
\begin{equation}
W_{ijp}^T\theta=\theta_0+\theta_1age+\theta_2gender+\theta_3edu+\theta_4member+\theta_5income,
\label{equ:24}
\end{equation}
where $i=1,2,3$ in equation (\ref{equ:22}) denote the three dimensions of health literacy respectively, and subject $p$'s coefficient alpha on different pairs of items are assumed to be equal: $\alpha_p\triangleq\alpha_{ijp}=1-{\rm exp}(W_{ijp}^T\theta)$. We implement our method to estimate the regression coefficients $\theta$ in equations (\ref{equ:24}), estimate the asymptotic variance of the estimators, construct their 95\% confidence intervals, and compute their p-values in the regression model based on the estimators' consistency. The result are presented in Table \ref{tab:7}.
\begin{table}[htbp]
\center
\begin{tabular}{ccc}
\hline
      & Estimate (95\% CI) & P-value \\
\hline
$\theta_0$ & -4.41 (-10.17, 1.35) & 0.129 \\
$age$ & 1.05 (-0.68, 2.77)  & 0.196 \\
$gender$ & 1.90 (-1.21, 5.01) & 0.195 \\
$edu$ & -4.05 (-15.59, 7.50) & 0.315 \\
$member$ & -2.65 (-8.05, 2.74) & 0.250 \\
$income$ & 0.559 (-1.83, 2.95) & 0.359 \\
\hline
\end{tabular}
\caption{\label{tab:7} Estimation of the coefficients in the third set of GEEs, with 95\% confidence interval and p-values.}
\end{table}

Results in Table \ref{tab:7} indicate there is no significant relationship between the internal consistency of the three dimensions of health literacy scores and the covariates of our interests. This fact demonstrates good homogeneity of coefficient alpha among the subjects. Since no variables are significantly related to the coefficient alpha in our model, we drop all of them from the third set of GEEs to simplify the design matrix and reduce the variance of the estimators, and then fit our model again to estimate the coefficient alpha that is actually homogeneous among the samples. The point estimation the coefficient alpha is $0.86$, which is in $(0.7,0.9)$, the scale recommended by (Tavakol and Dennich, 2011\cite{Tavakol2011Making}). And its 95\% confidence interval (CI) is $(0.72,1.00)$, estimated by delta method. Though the length of the interval is relatively high, the 95\% CI of the coefficient alpha is still within an acceptable scale. The p-values of the regression coefficients in the third set of GEEs shown in Table \ref{tab:7}, as well as estimation of the coefficient alpha with all of the covaraites dropped suggests good quality and reliability of the results in this health literacy test.

\section{Discussion}
To handle with heterogenicity of the samples, we define an individualized coefficient alpha for measurement of the test scores' internal consistency. We propose a three-set-of GEE method to model the newly defined coefficient alpha with covariates of our interests. It is a quasi-likelihood method, and one does not have to specify distribution of the responses. Missingness of a key covaraite is also considered in our method. Under mild assumptions, we can obtain consistent estimators for the regression coefficients in GEE, and the individualized coefficient alpha, which allows for interval estimation and hypothesis testing.

Simulation studies show that bias and mean squared errors of the estimators for our interested parameters is reasonable, when the settings of the simulation datasets are near the real dataset analyzed in this paper. As a results of the sample size's mildly increasing:
\begin{itemize}
\item The means of the estimators get closer to the true value, and their RMSE decrease strictly.
\item Type one error rate for the hypothesis testing on the negative variables approaches the given significant level (0.05).
\item Power for both testing on the regression coefficients and testing concerning our newly defined coefficient alpha increases in a reasonable speed.
\end{itemize}
These results demonstrate good convergence and asymptotic performance of our method. Application of our proposed method in Health Literacy data analysis also implies its potential and promising use in practice.

A limitation of our proposed method is that it would be of less power and be poor in interpretation if the variance of the parameters is high. Such high variance can result from relatively poor sample size or high variation of the covariates in practice. And our newly defined $\alpha_{ijp}$ is sensitive to individuals with outlying covariates. The sensitivity of the individualized coefficient alpha could make it difficult for us to analyze the overall internal consistency in our proposed framework. Thus, method for robust estimation of our defined individualized coefficient alpha is desired in future work to reduce the potentially significant influence of outliers.

\end{document}